# Gender diversity in research teams and citation impact

# in Economics and Management

Abdelghani Maddi[1] and Yves Gingras[2]

[1] abdelghani.maddi@hceres.fr
Observatoire des Sciences et Techniques, Hcéres, Rue Albert Einstein, Paris, 75013, France ; CEPN, UMR-CNRS 723, Université Paris 13, 99 Avenue Jean Baptiste Clément, 93430 Villetaneuse.

[2] gingras.yves@uqam.ca
Centre interuniversitaire de recherche sur la science et la technologie (CIRST), Université du Québec À Montréal, C.P. 8888, Suc Centre-Ville, Montréal, Qc, Canada, H3C 3P8.

## Abstract

The aim of this paper is twofold: (1) contribute to a better understanding of the place of women in Economics and Management disciplines by characterizing the difference in levels of scientific collaboration between men and women at the specialties' level; (2) Investigate the relationship between gender diversity and citation impact in Economics and Management. Our data, extracted from the Web of Science database, cover global production as indexed in 302 journals in Economics and 370 journals in Management, with respectively 153,667 and 163,567 articles published between 2008 and 2018. Results show that collaborative practices between men and women are quite different in Economics and Management. We also find that there is a positive and significant effect of gender diversity on the academic impact of publications. Mixed-gender publications (co-authored by men and women) receive more citations than non-mixed papers (written by same-gender author teams) or single-author publications. The effect is slightly stronger in Management. The regression analysis also indicates that there is, for both disciplines, a small negative effect on citations received if the corresponding author is a woman.

# 1 Introduction

Gender diversity (GD) in science is widely debated in the literature (Campbell *et al.* 2013a; Nielsen *et al.* 2018; Nielsen and Börjeson 2019). Although policy makers are convinced of the positive effect of diversity on team performance (Moore 2006; Page SE 2008; European Commission 2013; Lauring and Villesèche 2019a), results from empirical studies on the issue are mixed (Campbell *et al.* 2013a; Nielsen and Börjeson 2019).

The links between "performance" and GD are complex and include several dimensions. The literature has analyzed the question through several prisms like team cohesion (Webber and Donahue 2001), human capital and collective intelligence (Bozeman and Corley 2004), productivity (Abramo *et al.* 2009; Defazio *et al.* 2009; Lee and Bozeman 2016; Abramo *et al.* 2017), scientific discovery and innovation (Nielsen *et al.* 2018), citations impact (Lauring and Villesèche 2019b; Nielsen and Börjeson 2019), teams' performance based on peer evaluation (Lerback *et al.* 2020) or collaboration networks (Mcdowell *et al.* 2006; Kwiek and Roszka 2020).

This study focused on the link between GD, mixed collaborations and citations impact in the fields of Economics and Management. To do so, we used all publications in these two disciplines indexed in the WoS database between 2008 and 2018. This makes a sample of 153,667 published in 302 journals in Economics and 163,567 articles published in 370 journals in Management. A regression analysis has been performed to highlight the relationship between the number of citations received and the gender composition of research articles in the two disciplines. More precisely, we constructed five groups of articles according to the sex of authors: (1) articles co-signed by men and women, (2) by women only (at least two), (3) by men only (at least two), (4) authored by single woman and (5) by single men. Using Tobit regression, we established a hierarchical order between the different types of publication in terms of citations received. Several control variables were used, such as the number of authors, number of countries, journal Impact factor and the research area.

The article is organized as follow. In section 2, we present an overview of the literature dealing with the link between gender and citation impact. Section 3 describes the data and method used. Section 4 presents descriptive statistics on collaboration on gendered collaborations, and the results of a Tobit regression.

# 2 Literature review on gender and citation impact

The literature offers a broad spectrum of analysis on gender and academic impact. The issue has been addressed from many perspectives. Different studies have analyzed the effect of positions of authors according to gender and citations rates (Larivière *et al.* 2013; Thelwall and Sud 2020), self-citation practices according to gender (King *et al.* 2017; Mishra *et al.* 2018; Azoulay and Lynn 2020) and gender differences according to the age of the authors (Frandsen *et al.* 2020; Kwiek and Roszka 2020).

Using a large sample from the WoS database between 2008 and 2012, consisting of 5,483,841 publications involving 27,329,915 authors, Larivière *et al.* (2013) found that publications with women in first or last author positions receive fewer citations than those with men in the same positions. Insofar as international collaboration increases the visibility of articles and the number of citations received (Rigby and Edler 2005), this citation disadvantage is explained, in part, by the fact that women have less international collaboration than their male counterparts. Several studies on different countries confirm the lower level of international collaboration of women researchers. This is for example the case in Italy (Abramo *et al.* 2013), Poland (Kwiek and Roszka 2020), Denmark (Nielsen 2016) and Norway (Aksnes *et al.* 2019). In the case of India, Paswan and Singh (2020) have shown that although articles by male authors have a citation advantage, Indian women researchers tend to have more international collaboration. On this issue of international collaboration, it is important to emphasize that the distributions of men and women by discipline are different (Aksnes *et al.* 2019; Paswan and Singh 2020). As pointed out by Aksnes *et al.* (2019), the concentration of women is higher in fields where international collaboration is generally lower.

Several studies have shown that gender gaps still persist in favor of men both in terms of productivity and scientific visibility (Abramo *et al.* 2009; Duch *et al.* 2012; West *et al.* 2013; Beaudry and Larivière 2016; Budrikis 2020; Odic and Wojcik 2020). In the case of Psychology, (Odic and Wojcik 2020) have shown a significant effect of gender in favor of men who receive on average 19% more citations. It has also been observed that the greater presence of men in highly cited journals could be an important factor explaining these differences in citations (Nielsen 2016).

Maddi *et al.* (2019) have shown that the number of citations received in Economics and Management is negatively affected by the presence of women as co-authors. The period considered was between 2008 and 2015 with approximately 170,000 publications. Both disciplines were considered as a whole, that is without taking account of the different specialties of these disciplines. Given that the level of citations can vary considerably depending on the specialties, the citation advantage of men could be due to the distribution of men and women within these different specialties. For example, citation levels are very high in "Environment, agriculture, natural resources, and energy" area and lower in "Accounting and auditing". As the proportion of men and women also varies depending on the specialty, it is important to consider this variable to better measure the specific effect of gender on citation rates (Maddi *et al.* 2019).

Based on a sample of 437,787 publications authored by 4,292 faculty members from the top universities in the United States, Duch *et al.* (2012) showed that gender differences in impact rate are discipline specific. In addition, the authors argue that the lower performance of women in certain disciplines is linked to the lack of resources to which they have access. Duch *et al.* (2012) emphasize the importance of gender diversity within disciplines to overcome gender differences in productivity and citation impact.

Another factor that could be at the origin of the differences in impact observed between the two sexes are self-citations. Using a sample of 1.5 million articles from JSTOR database between 1779 and 2011, King *et al.* (2017) have shown that nearly 10% of references are authors self-citations. They have also shown that men cited their own papers 56% more often than women did. In the last two decades of data, men self-cited themselves 70% more than women did. Other studies have, however, arrived at contradictory results. This is the case of Mishra *et al.* (2018) on biomedical articles in PubMed database, and that of Azoulay and Lynn (2020) who examined 36 cohorts of life scientists (1970–2005). These studies have shown that men and women self-cite at the same rate. We therefore cannot generalize (King *et al.* 2017)'s result, which does not seem to be valid in the life sciences. In a sample of 3,923 Danish researchers (7,820 publications), Nielsen (2016) also found no specificities linked to gender in terms of self-citation practices. In political science, a recent study has shown that self-citation practices depend on several factors like scientific collaboration and previous experience of authors (Dion *et al.* 2020). Overall these authors have shown that men are more likely to cite their previous work than women.

The issue of gender differences in citations impact was also been looked at in terms of age and academic career. Based on data from the population of university professors in the province of Quebec (Canada), Larivière *et al.* (2011) showed that passed the age of 38, the disadvantage of women in terms of productivity and impact is higher. The authors provide several possible explanations, such as the smaller collaborative networks of women, motherhood and the division of labor that accompanies it, difficulty accessing resources and funding. These difficulties confronting women impact their productivity from the start of their career (Nielsen 2018). Beaudry and Larivière (2016) have shown that, in natural sciences and engineering, for equal amounts of funding and publications, women are equally cited.

In a study based on a sample of 25,463 Polish university professors from 85 universities, grouped into 27 disciplines (Scopus database), Kwiek and Roszka (2020) showed that in addition to international collaboration, the disparities in citation impact between the two sexes are also linked to age. Until about 40 years old, the differences are marginal and then they start to grow in favor of men. In a broad analysis of science, technology, engineering, and mathematics (STEM) fields, Huang *et al.* (2020) provided a comprehensive picture of gender differences in scientific performances with a focus on the academic careers of researchers. The authors reconstructed the full publication history of over 1.5 million authors, identified by gender, from 83 countries and whose publishing careers was between 1955 and 2010. The results showed that, overall, men and women publish during their career a comparable number of articles with an equivalent average impact. However, the increase in the number of women during the 60 years of observation paradoxically widened the differences between men and women. The authors concluded that the length of the publishing career and the dropout rate could explain the differences observed between the two sexes.

Contrary to the studies presented above, some papers find an advantage of citations in favor of women. Using a sample of nearly 27,000 publications and more than 65,000 authors, Nielsen (2017) analyzed the differences in academic impact in Management sciences by gender and concluded that women have a slightly greater impact than men, while remaining cautious about the representativeness of their sample and the possibility of generalization. Similarly, women have a larger share in the decile of the most cited publications in Management. In a more recent study, based on a sample of six million articles from the Scopus database between 1996 and 2018 (countries: Australia, Canada, Ireland, Jamaica, New Zealand, United Kingdom and United States), Thelwall (2020a) showed the existence of a small advantage of citations in favor of women over the years and for all countries studied except the United States. The female citation advantage is more significant for Australia and the UK. In a similar study, examining publications from the same seven countries (Thelwall 2020a), Thelwall (2020b) also showed the existence of a female citation advantage in dominant author positions.

By contrast, Nielsen (2016) using a sample of Danish researcher, Thelwall and Sud (2020) analyzing six million articles published between 1996 and 2012 across up to 331 fields, and Frandsen *et al.* (2020) studying health science researchers, found no effect (positive or negative) of gender on citation impact. Likewise, Leimu and Koricheva (2005a) showed that there is no empirical evidence for an effect of first author position on the citation rate in the case of ecological papers.

Regarding the relationship between GD and academic impact, previous studies on the issue have shown that, overall, GD has no or only a moderate effect on citations impact. Results differ according to the variables studies as well as the disciplines. In environmental sciences, Campbell *et al.* (2013b) showed that articles co-signed by men and women receive, on average, 34% more citations than those written by teams of same-sex authors. However, the positive effect of GD diminishes as the share of women increases in heterogeneous gender teams. Based on 25,000 papers in Management field, Nielsen and Börjeson (2019) found no effect of GD on citation impact. However, they argued that GD could lead to expanded research programs. In contrast, in political science, sociology, and economics, Dion *et al.* (2018) have shown that the more mixed the research teams, the lower the gender citations gap. GD therefore appears to be a solution to reduce the inequalities in citations received between women and men (known as the "Matilda effect": see Rossiter (1993)). In a recent article, Lerback *et al.* (2020) showed that the GD of teams positively affects the acceptance rate of scientific publications; the rate is 4.5 % higher for publications including both sexes. However, citations are slightly lower for publications co-signed by women and men, compared to mono-gender publications.

From this long literature review, we can stress the great diversity of the results obtained according to the samples and the variables studied. Whereas some papers suggest an advantage of citations in favor of men, others find the opposite effect or no effect at all. Overall, once the controls on disciplines, equal opportunities, access to resources and grants, etc. are taken into account, it is not easy to conclude that there is a systematic and universal advantage in favor of men (West *et al.* 2013). Furthermore, the literature also showed the importance of experience, age and position in academic career in determining citation impact. In short, it is possible that specific effects do vary according to disciplines. Here we will focus on Economics and Management as they are closely related but also different in terms of methods and collaboration practices.

## 3  Data and method

### a.  Data

***Database***

The data have been extracted from the Web of Science (WoS) Core Collection database (SCIE, SSCI, A&HCI). Only "articles" are taken into account in the study, other types of documents (such as editorials, letters, etc.) are excluded from the analysis because they are generally not original contributions to scholarly knowledge (Moed 1996).

The WoS database gives us information on authors, titles of publications, institutional address (institution, city, country, etc.), citations number received by publications, journal title. We have added variables like the gender of the authors, the number of authors per publication, the number of countries per publication, impact factor of journals, country of journals (publisher) and the research area of journals.

*Name gender assignment*

In this study, we analyze whether GD affects the number of citations received by scientific publications. To represent gender in scientific publications, we used the proxy of the proportion of women per publication. The authors' gender is assigned based on the methodology presented in Larivière *et al.* (2013), which uses the author's first name to assign gender. For each of the articles, we calculated the proportion of authors belonging to each gender, using as denominator the sum of the authors to whom we assigned a gender. For example, an article with five authors, including two women, two men, and one unknown, was assigned a proportion of female authors of 0.5, leaving unknown cases out of the calculation. For an article co-signed by men only, the proportion is 0, and for an article whose all authors are women the proportion is 1. The values between 0 and 1 represent articles co-authored by both men and women (mixed-gender publications). The higher the number of women per publication, the more the proportion is closer to 1.

We summarize here the main steps of determining the sex of the authors. For more details, see Larivière *et al.* (2013).

The gender-detection method matches the author list with universal and country-specific name lists. The list of author names requires pre-processing. To be able to do the matching, the pre-processing consists in removing the special characters in the names, identifying if they are initials, separating the composed names by spaces and replacing hyphens by a space ("Jean-Marc" become "Jean Marc").

Once the pre-processing has been carried out, the author's given names are matched with the following lists (for lists description see: Larivière *et al.* (2013)) in the same order below:

1. US Census;
2. WikiName;
3. Wikipedia;
4. France and Quebec list;
5. Other country-specific lists.

It should be noted that the gender of the authors is identified for each article separately. In other words, we did not carry out matching within the corpus to measure the number of distinct authors or the number of publications per author. This choice is determined by our research question, insofar as we analyze the impact of the gender composition of the articles on the citations received. Table 1 provides name disambiguation and gender identification statistics for the two disciplines (Economics and Management). Out of the total number of authors, we identified the gender for about 83% of them.

**Table 1: proportion of full names and of given names assigned a gender**

| Type | | % | |
|---|---|---|---|
| | | Economics | Management |
| Assigned | Men | 63,4% | 57,7% |
| | Women | 20,3% | 25,1% |
| | *Total assigned* | *83,7%* | *82,8%* |
| Not assigned | Initials | 3,9% | 3,8% |
| | Unisex | 2,8% | 2,8% |
| | Unknown | 9,6% | 10,6% |
| | *Total not assigned* | *16,3%* | *17,2%* |

The proportion of women is about the same in Economics (25.3% but they contribute for 32% of the publications) and Management (25.1% for 26% of publications). These distributions are near the average of the global distribution of women researchers, which is estimated to be around 30% (UNESCO, 2018).

***Disciplinary classification: research area of journals***

The finest disciplinary classification of WoS database consists of 254 specialties. Economy and Management are not subdivided into specialties, despite the fact that there is a great diversity within these two disciplines. To make an analysis at the level of the research area, a more detailed classification is needed. Therefore, we used the 2019 CNRS (French National Centre for Scientific Research) classification of journals into "specialties" of Economics and Management based on peer evaluation. We thus matched the CNRS list and that of the WoS using the journals names (after cleaning and homogenization). We relied on the expertise of one of us (M.A.) to assign the journals that did not match to one of the 23 specialties of CNRS list. To do so, the information provided on journals websites is used to assign unclassified journals (260 out of 672) to one of the 23 research areas presented in Table 6. The distribution presented in Table 6 provides a global view of the scientific production in the WoS and does not of course include all journals but only the most cited. As it can be seen in the figure 1, the CNRS classified, in 2019, 1674 journals in both disciplines, while the WoS indexes only 412 of them (24.6%).

**Figure 1: overlap between journals indexed in 2019 CNRS classification and WoS database**

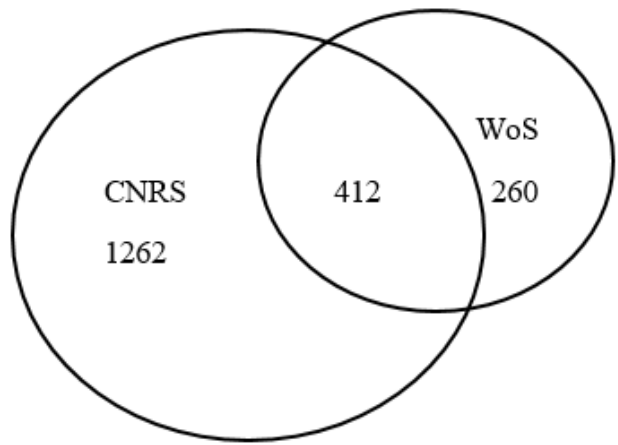

### b. Method

A regression analysis was performed to study the relationship between GD and citation impact. The choice of method depends essentially on the nature of the dependent variable. In our case, we use the citations received by articles as proxy for academic impact.

For our regression, we used Normalized Citation Scores (NCS). Recall that the NCS of a given article is calculated by dividing its number of citations received by the average number of citations in the same disciplines and the same year. Indeed, we uses a Tobit regression rather than a negative binomial, since NCS are not account data. The advantage of using the NCS indicator is that it corrects the year effect. The dependent variable is the number of citations received by each publication during the period 2008-2018. To retain the zeros, we have added 1 to the NCS before making the logarithmic transformation. Log (NCS) is a continuous variable with a lower boundary at zero and an upper boundary at infinity. Thus, a left censored Tobit regression model is used (McDonald and Moffitt 1980) to account for the disproportionate number of observations with zero values, because a significant proportion of the observations in our sample are zeros. Tobit regression avoid inconsistent estimates from Ordinary Least Square (OLS) regression.

*Independent variables*

To represent GD we constructed five dummy variables to denote articles published only by a (1) single man (M_single) or by a (2) single woman (W_single), articles involving (3) only men (at least two) (M-M) or (4) only women (at least two) (W-W) and articles involving (5) both men and women (at least one man and one woman) (M-W). Insofar as it denotes GD, the variable representing articles co-authored by men and women is used as a reference modality in the regressions.

Table 2 shows the distribution of publications according to the five types of publication. We observe a much larger proportion of collaboration between men and women in Management that in Economics. This difference also appears in the fact that publications involving a single male author are more important in Economics than in Management. Economics has thus more than half (52%) of its papers written by men only whereas that proportion is only in 40% in Management.

In both disciplines, only about 5% are written by women alone and 10% in collaboration with other women.

**Table 2: distribution of publications by gender of authors**

|          | Economics | | Management | |
|----------|-----------|------|------------|------|
|          | #         | %    | #          | %    |
| W_single | 7579      | 5%   | 6274       | 4%   |
| M_single | 34395     | 22%  | 20250      | 12%  |
| W-W      | 14643     | 10%  | 17018      | 10%  |
| M-M      | 45702     | 30%  | 45849      | 28%  |
| M-W      | 51348     | 33%  | 74176      | 45%  |
| Sum      | 153667    | 100% | 163567     | 100% |

*Control variables*

A number of control variables are included in our analysis. The choice of control variables comes from the literature that shows that they are potentially associated with the number of citations received by publications (Larivière *et al.* 2011; Beaudry and Larivière 2016). Judge *et al.* (2007) analysis of the citation determinants of articles published in the top 21 Management journals showed that the main factor in the visibility of an article is the journal in which it is published. Harzing (2016) showed that it is rather the topic studied by the article, as well as the profile of the author, that influences the visibility of publications in Management. Starbuck (2005) and Singh *et al.* (2007) concluded that the evaluation of research articles based solely on the impact of journals provides erroneous results as to the "quality" of publications, given intra-review variability. In addition, there is an abundant literature showing that citation scores are strongly linked to scientific collaboration (Smart and Bayer 1986; Van Raan 1998; Hara *et al.* 2003; Leimu and Koricheva 2005b; Franceschet and Costantini 2010; Bote *et al.* 2013; Larivière *et al.* 2015).

First, we control for the number of authors (Nbr_Authors) and the number of countries by publication, a proxy of international collaboration (Internat_collab). Second, we have controlled for the geographic origin of journals by building two dummy variables. The country of publisher was used as proxy of country of journals. US_Journal takes the value 1 if journal is American. *EU_Journal* takes the value 1 if journal is European. The non-American and non-European journals are the reference variables. Third, we control the impact of journals in which articles are published using the 2 years Impact Factor of the journal. And finally, we controlled by the specialties using dummies variables. We have included the five main specialties (research areas) of each of the two disciplines.

Using exactly the same variables, two distinct regressions were performed for both disciplines. The aim is to analyze the differences between Economics and Management regarding the impact of GD on the citation score. The Table 3 summarizes variables of model.

**Table 3: Dependent, explicative and control variables in the regression**

| Dependent variable | |
|---|---|
| $\text{Log}(NCS)_i$ | Log transformed of NCS (Normalized Citations Score) by publication $i$. |
| **Explicative variables** | |
| $\text{M\_single}_i$ | Dummy variable that takes 1 if the publication contains a single male author. |
| $\text{W\_single}_i$ | Dummy variable that takes 1 if the publication contains a single female author. |
| $M - M_i$ | Dummy variable that takes 1 if the publication contains authored only by men (at least two). |
| $W - W_i$ | Dummy variable that takes 1 if the publication contains authored only by women (at least two). |
| $M - W_i$ | Dummy variable that takes 1 if the publication involves both at least one man and one woman. |
| **Control variables** | |
| $Corr\_aut\_w$ | Dummy variable indicating the fact that corresponding author is woman. |
| $Nbr\_Authors_i$ | Number of authors by publication. |
| $Internat\_collab_i$ | International collaboration measured by the number of countries by publication. |
| $US\_Journal_i$ | Dummy variable indicating the fact that the publisher of journal is American. It equal to 1 if it is. |
| $EU\_Journal_i$ | Dummy variable indicating the fact that the publisher of journal is European. It equal to 1 if it is. |
| $IF\_2$ | 2 years Journal Impact Factor. |
| Dummies variables by specialty (research area): | |
| $GEN$ | General Economics, general Management |
| $AgrEnEnv$ | Environment, Agriculture, Natural resources, Energy |
| $Macro$ | Macroeconomics, international and monetary Economics |
| $DevTrans$ | Development and transition Economics |
| $ThEco$ | Economic theory, game and decision theory and experimental Economics |
| $SI$ | Management information systems |
| $Fin$ | Finance and insurance |
| $GRH$ | Human resources Management |
| $MKG$ | Marketing |

Before interpreting the coefficients, we have verified that all the the variance inflation factors (VIF) scores are much lower than five. This suggests the absence of multi-collinearity (James *et al.* 2017; Bruce *et al.* 2020).

## 4   Descriptive statistics

Our data, extracted from the WoS database, cover global production as indexed in 302 journals in Economics and 370 journals in Management, with respectively 153,667 and 163,567 articles published between 2008 and 2018 (Table 5). The respective contribution of women in both disciplines is 28% and 32%.

**Table 5: number of publications, journals and proportion of women by discipline**

|  | Economics | Management | All |
|---|---|---|---|
| # publications | 153 667 | 163 567 | 317 234 |
| # journals | 302 | 370 | 672 |
| % average fraction of women on papers | 28% | 32% | 30% |

**Table 6: number and proportion of publications, by specialty by discipline and Sex**

| Research area code | Research Area | Economics | | | | | | Management | | | | | |
|---|---|---|---|---|---|---|---|---|---|---|---|---|---|
| | | # | % | % average fraction of women on papers | % M-W | % W only | % M only | # | % | % average fraction of women on papers | % M-W | % W only | % M only |
| GEN | General Economics, general Management | 46 499 | 30,3 | 27,7 | 29% | 15% | 56% | 21 517 | 13,2 | 32,7 | 44% | 15% | 41% |
| AgrEnEnv | Environment, Agriculture, Natural resources, Energy | 29 004 | 18,9 | 35,5 | 46% | 11% | 43% | 1 157 | 0,7 | 34,7 | 47% | 16% | 37% |
| Macro | Macroeconomics, international and monetary Economics | 16204 | 10,5 | 35,2 | 29% | 16% | 55% | 980 | 0,6 | 26,7 | 33% | 13% | 54% |
| DevTrans | Development and transition Economics | 10096 | 6,6 | 22,0 | 36% | 20% | 44% | 922 | 0,6 | 37,7 | 52% | 16% | 31% |
| ThEco | Economic theory, game and decision theory and experimental Economics | 9 742 | 6,3 | 26,4 | 24% | 15% | 61% | 2 450 | 1,5 | 31,5 | 49% | 13% | 38% |
| EcoPub | Public Economics and public choice | 8 769 | 5,7 | 27,6 | 27% | 15% | 58% | 1 216 | 0,7 | 31,5 | 36% | 17% | 47% |
| Metrie | Econometrics | 7 026 | 4,6 | 26,7 | 31% | 16% | 53% | 1 746 | 1,1 | 25,8 | 35% | 12% | 54% |
| OrgInd | Industrial organization | 4968 | 3,2 | 35,0 | 32% | 12% | 56% | 52 | 0,0 | 29,5 | 23% | 19% | 58% |
| Fin | Finance and insurance | 4 305 | 2,8 | 21,9 | 33% | 14% | 53% | 19 396 | 11,9 | 28,3 | 36% | 13% | 51% |
| TravPop | Labor and population Economics | 4 268 | 2,8 | 32,1 | 36% | 16% | 48% | 1 376 | 0,8 | 36,3 | 40% | 20% | 40% |
| LOG | Production and operations Management | 3 174 | 2,1 | 32,0 | 43% | 15% | 42% | 8 274 | 5,1 | 32,0 | 48% | 13% | 39% |
| Spatiale | Urban, spatial and regional Economics, transportation and tourism | 2 239 | 1,5 | 28,5 | 40% | 11% | 49% | 3 268 | 2,0 | 39,6 | 51% | 19% | 30% |
| SI | Management information systems | 1 477 | 1,0 | 29,1 | 47% | 11% | 42% | 20 795 | 12,7 | 31,6 | 45% | 14% | 40% |
| EcoDroit | Law and Economics | 1452 | 0,9 | 34,6 | 23% | 12% | 65% | 172 | 0,1 | 22,1 | 19% | 15% | 66% |
| Innov | Innovation and entrepreneurship | 1 403 | 0,9 | 25,5 | 48% | 13% | 40% | 9 967 | 6,1 | 32,3 | 47% | 14% | 39% |
| MKG | Marketing | 1 127 | 0,7 | 30,0 | 36% | 20% | 44% | 17 581 | 10,7 | 36,2 | 54% | 15% | 31% |
| HPEA | History of economic thought, economic and business history, methodology | 871 | 0,6 | 27,0 | 11% | 17% | 72% | 0 | 0,0 | - | - | - | - |
| StratInt | Business strategy and international Management | 424 | 0,3 | 27,3 | 23% | 17% | 60% | 7 041 | 4,3 | 32,7 | 47% | 14% | 39% |
| GRH | Human resources Management | 345 | 0,2 | 25,0 | 40% | 20% | 40% | 17 797 | 10,9 | 34,3 | 53% | 14% | 33% |
| CPT | Accounting and auditing | 274 | 0,2 | 31,2 | 40% | 19% | 41% | 7 853 | 4,8 | 32,3 | 45% | 14% | 41% |
| RO | Operations research | 0 | 0,0 | - | - | - | - | 13 927 | 8,5 | 26,8 | 39% | 12% | 50% |
| SANT | Health Economics and Management | 0 | 0,0 | - | - | - | - | 122 | 0,1 | 31,1 | 48% | 12% | 40% |
| ThO | Organization studies | 0 | 0,0 | - | - | - | - | 5 958 | 3,6 | 35,0 | 43% | 17% | 39% |
| **All** | | **153 667** | **100 %** | **28 %** | **33%** | **14%** | | **163 567** | **100 %** | **32 %** | **45%** | **14%** | **39%** |

We observe that the distribution is contrasted according to specialty and discipline (table 6). In Economics, nearly 60% of publications are concentrated in three specialties: General Economics, General Management (30%), Environment, Agriculture, Natural Resources, Energy (19%) and Macroeconomics, International and Monetary Economics (10%). In Management, the distribution is less concentrated, with specialties such as General Economics, General Management (13%), Management Information Systems (13%), Finance and Insurance (12%), Human Resources Management (11%) and Marketing (11%) having the highest proportion of publications (>10%).

## 5   Evolution of Collaboration practices in Economics and Management

At the global level, scientific collaboration, measured by the number of authors per article, is relatively stronger in Management than in Economics. The proportion of articles with at least two authors is 87% in Management against 77% in Economics (see Figure 2). This is a global average and the results vary somewhat by country. Over the decade 2008-2018, he collaboration rate has increased significantly in both disciplines. The share of publications signed by a single author decreased by 10% in Economics and 7% in Management, between 2008-2010 and 2016-2018.

**Figure 2: Number of authors per publication in Economics and Management**

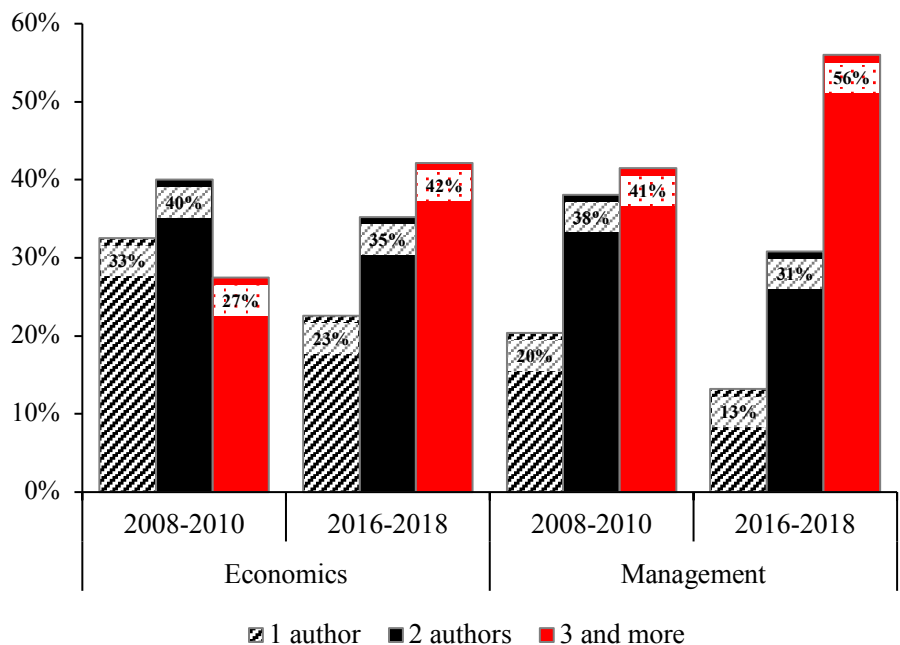

Figure 3 shows the evolution of male-female collaboration over the last ten years. We note that the proportion of single-sex collaboration as well as single female authors remain stable over the period. On the other hand, publications with a single male author have fallen sharply in favor of publications with male-female collaboration (especially in Economics). This shows a significant shift in collaborative practices between the two sexes. Men who tended to publish alone are more and more inclined to collaborate with other authors including at least one woman.

In 2018, the rate of men-women collaboration in Economics reached 40% and 55% in Management. In Economics, the proportion of publications by single men declined from 27% in 2008 to 17% in 2018 in favor of men-women papers.

**Figure 3: Collaboration between men and women in Economics and Management**

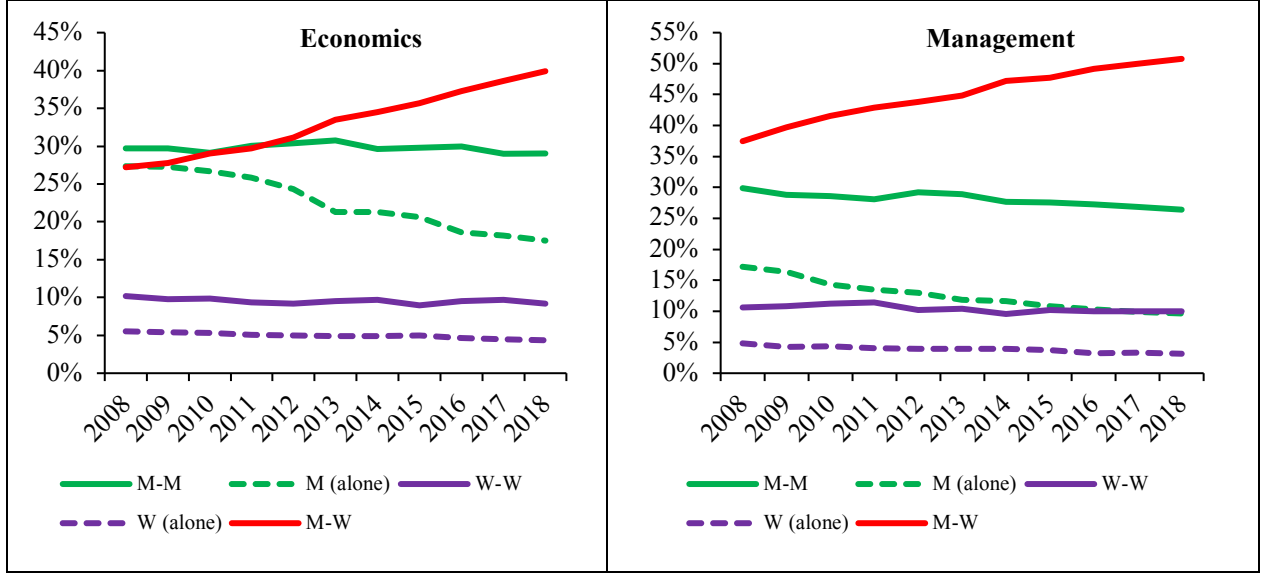

Figures 4 and 5 shows that male-female collaboration practices differ significantly among the specialties. In Economics, Development and transition Economics is the specialty with the highest share of publications including women (20% of publications are signed by women only, and 36% involve at least one woman). In contrast, 72% of "History of economic thought, economic and business history, methodology" publications are signed by men only (without women collaboration). The rate of men-women collaboration in this specialty is 11%. Similarly, in Management, the Development and transition Economics specialty has the largest share of publications involving women, nearly 70%. This share is comparable to that of Marketing. The specialties with the highest proportion of publications including only men are Law and Economics (70%) and Industrial organization (60%). We also note that, overall, the percentage of collaboration between men and women is significantly higher in Management than in Economics in almost all specialties.

It would be interesting to look further into the question of the collaborative strategy based on gender in the case of Economics and Management. Studies on the issue show that there are several factors that can guide the choice of collaboration (Bozeman and Gaughan 2011). Bozeman and Corley (2004) showed that women scientists have a 12% higher percentage of female collaborators than men (36% against 24%). However, Bozeman and Corley (2004) also shows that there are big differences according to rank, as non-tenure track females have on average of 84% of their collaborations with women.

**Figure 4: Distribution of publications by specialty and by collaboration type (Economics)**

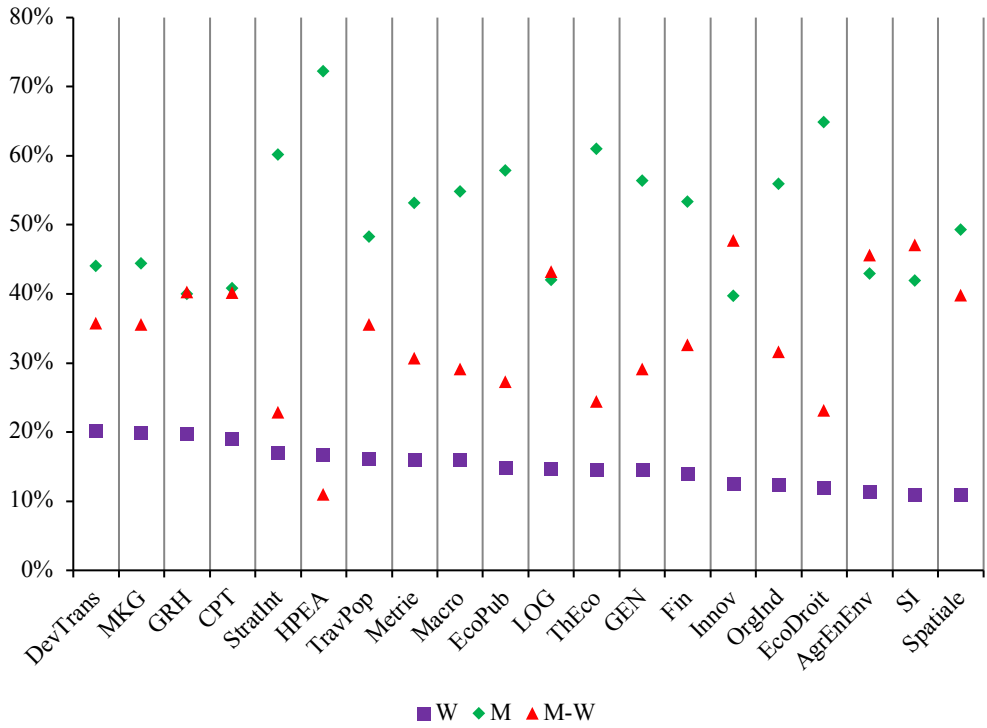

**Figure 5: Distribution of publications by specialty and by collaboration type (Management)**

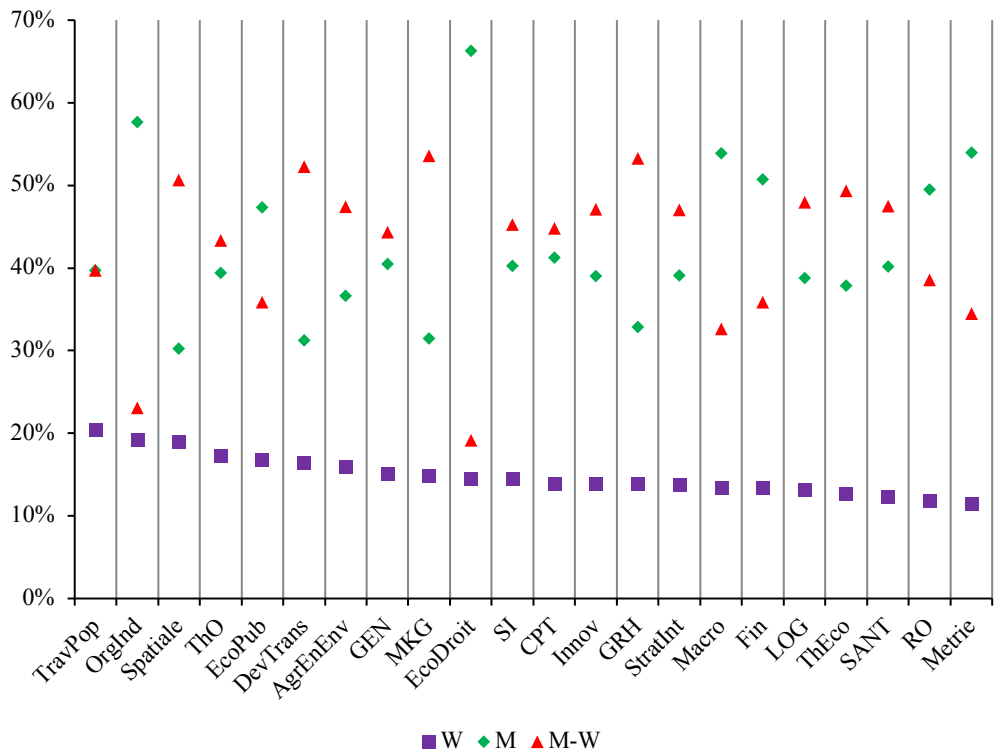

Although the proportion of articles written in collaboration is greater in Management, the distribution between national and international co-publications by gender is similar in both disciplines. This is true for co-publications that include only women, only men, or both (Table 7). It should be noted, however, that men-women collaborations have a higher proportion of international co-publications than men only or women only. The share of international co-publications in the two disciplines is about 30%.

**Table 7: number and proportion of international collaboration by discipline**

| International Collaboration | Economics | | Management | |
|---|---|---|---|---|
| M | 19 179 | 23,9% | 16 212 | 24,5% |
| W | 5 979 | 26,9% | 5 518 | 23,7% |
| M-W | 20 751 | 40,4% | 27 859 | 37,6% |
| **All** | **45 909** | **29,9%** | **49 589** | **30,3%** |

Figure 6 shows that publications without collaboration are the least cited, regardless of the author's gender. They are 20 to 30% less cited than the world average (defined as 1.0). In contrast, publications with male-female collaboration are the most cited in the two disciplines, with a normalized citation score 10-20% higher than the world average. In addition, Figure 6 shows that co-publications involving only men are generally more cited than those including only women in both disciplines.

**Figure 6: Normalized citations score by men-women collaboration type**

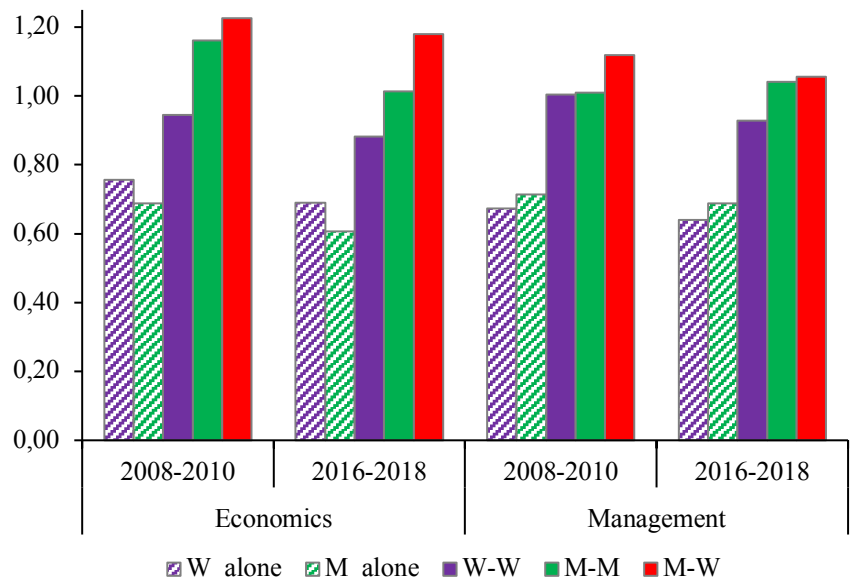

## 6    Results of regression analysis

We distinguish four types of independent variables: sociological (authors' number and international collaboration), geographical (the fact that the journal is American or European), bibliometric (impact of journals measured by Impact Factor) and disciplinary (23 specialties).

**Table 8: Tobit maximum likelihood estimation**

| | Variables | Economics | | Management | |
|---|---|---|---|---|---|
| | | Coefficient | Pr (>|z|) | Coefficient | Pr (>|z|) |
| Gender | W_single | -0.04*** | 0.0000 | -0.07*** | < 2e-16 |
| | M_single | -0.08*** | < 2e-16 | -0.09*** | < 2e-16 |
| | W-W | -0.02*** | 0.0001 | -0.03*** | 0.0000 |
| | M-M | -0.01 | 0.1325 | -0.02*** | 0.0000 |
| | M-W (ref.) | - | - | - | - |
| | Corr_aut_w | -0.02*** | 0.0000 | -0.01*** | 0.0002 |
| Collaboration | Nb_Authors | 0.024*** | < 2e-16 | 0.003** | 0.0555 |
| | Nb_countries | 0.05*** | < 2e-16 | 0.04*** | < 2e-16 |
| Journal | ln_IF_2 | 0.75*** | < 2e-16 | 0.60*** | < 2e-16 |
| | US_Journal | 0.10*** | < 2e-16 | 0.06*** | < 2e-16 |
| | UE_Journal | 0.06*** | < 2e-16 | 0.03*** | < 2e-16 |
| Disciplines | GEN | -0.09*** | < 2e-16 | -0.07*** | < 2e-16 |
| | MKG | -0.10*** | 0.0000 | 0,00 | 0.2957 |
| | GRH | -0.01 | 0.8370 | 0.02*** | 0.0000 |
| | DevTrans | -0.01 | 0.1539 | -0.03* | 0.077377 |
| | ThEco | -0.1*** | < 2e-16 | -0.12*** | < 2e-16 |
| | EcoPub | -0.04*** | 0.0000 | -0.12*** | 0.0000 |
| | Metrie | -0.06*** | 0.0000 | -0.09*** | 0.0000 |
| | OrgInd | -0.03*** | 0.0022 | -0.16** | 0.0493 |
| | Fin | -0.04*** | 0.0000 | -0.10*** | < 2e-16 |
| | TravPop | 0,00 | 0.6110 | -0.14*** | < 2e-16 |
| Model statistics | Wald-statistic | 4.982e+04 | < 2.22e-16 | 3.54e+04 | < 2.22e-16 |
| | Log-likelihood | -1.301e+05 | | -1.302e+05 | |

*** significant at 1% / ** significant at 5% / * significant at 10%.

Table 8 shows four important results that we can summarize as follows:

- The Tobit regression results indicate the existence of a moderate positive and significant (except for the case of co-publications in Economics involving only men) effect of GD on the number of citations received. The sign of all the coefficients of the authors' gender variables is negative in both regressions (for Economics and Management). This means that the impact of publications involving both men and women (reference modality) is higher. The GD effect is slightly stronger in Management.

Although publications with GD are the most cited in both disciplines, the regressions show the existence of a hierarchy according to the collaboration and the gender of authors. The order is the same for both disciplines. Thus, publications with a single male author are the least cited in both disciplines, followed by publications with a single female author. Publications involving several authors of the same sex are relatively more cited with similar coefficients.

- On the importance of collaboration, regression results show that, for both Economics and Management, citations are positively and significantly shaped by the number of authors and the number countries involved in a publication. This result is consistent with the literature on the issue (Smart and Bayer 1986; Van Raan 1998; Hara *et al.* 2003; Leimu and Koricheva 2005b; Franceschet and Costantini 2010; Bote *et al.* 2013; Larivière *et al.* 2015).

- For both disciplines, the country of journal has a significant impact on citations. Thus, the fact that the journal in which the article is published is American or European increases the number of citations, which is not the case for journals published in any other country (that is outside US and EU). However, citations received are higher if the journal is American than if it is European.

- As could be expected, the academic impact of the journal is the variable that most affects the number of citations received by articles. The more the journal in which the article is published has a high impact factor and is thus more visible, the higher the number of citations of that article. Thus, the highest regression coefficients are recorded for the journal's "impact factor" variable, in both disciplines.

## 7  Discussion and conclusion

In this paper, we have investigated the relationship between gender diversity and citations received by academic papers in Economics and Management. Our results show the existence of a positive and moderate effect of GD on the citations received. This is consistent with some studies on the topic like those of Campbell *et al.* (2013b) and Dion *et al.* (2018). They are however different from those of Nielsen and Börjeson (2019) who found no effect of GD on citations. Our results also indicate that publications involving a single male author are the least cited in the two disciplines, followed by publications with a single female author. Publications involving multiple authors of the same sex come second in terms of impact after those with GD.  All this suggest that diversity do indeed stimulate better reflections that may lead to more original results.

Our data also highlight for the first time that the practice of collaboration between different genders is quite different according to specialties in Economics and in Management. Explanations of such practices would require an in-depth, interview-based qualitative study, but highlighting such differences in collaborative practices is in itself an important result. Our data also shows that publications with a single male author have fallen sharply in favor of publications with male-female collaboration, especially in Economics. Thus, men who tended to publish alone are more and more inclined to collaborate with other authors including at least one woman.

Our results also indicate that the visibility of research articles in Economics and Management is closely linked to the visibility of the journal in which they are published. This was to be expected because we know that there is a Matthew effect related to the journal impact factor (Larivière and Gingras 2010). Another important result in the current context of bibliometric evaluation of research is the weight of American journals in the visibility of research articles both in Economics and Management. Indeed, if the journal is American, the citations to the articles will nearly double compared to a European journal. It is likely that the important role of US journals (as the country of publication of the journal) in determining publication visibility as measured by citations is related to the fact that the WoS database (just like that of SCOPUS by Elsevier) has a strong Anglo-Saxon bias (Gingras and Khelfaoui 2018). It remains true, however, that researchers' evaluations are, in practice, based on these databases. Our results are therefore all the more important as they may in turn influence the future publication practices of scholars in order to improve their "score" of citations. It should also be noted that such a move toward American journals instead of European ones in Economics and Management, could also affect the choice of research topics and thus diminish knowledge production on European economic and management realities.

**Acknowledgments:** the authors thank Vincent Larivière for his help in the production of data including the gender assignment to authors.